\documentclass[reprint,aps,prl,twocolumn,superscriptaddress,showpacs,floatfix]{revtex4-1}

\usepackage{graphicx,enumerate}
\usepackage{amsmath,epsfig}

\newcommand{\be}{\begin{equation}}
\newcommand{\ee}{\end{equation}}
\newcommand{\ba}{\begin{eqnarray}}
\newcommand{\ea}{\end{eqnarray}}

\begin{document}

\title{Testing General Relativity with the Shadow Size of Sgr~A$^\ast$}

\author{Tim Johannsen}
\affiliation{Perimeter Institute for Theoretical Physics, Waterloo, Ontario N2L 2Y5, Canada}
\affiliation{Department of Physics and Astronomy, University of Waterloo, Waterloo, Ontario N2L 3G1, Canada}

\author{Avery E. Broderick}
\affiliation{Perimeter Institute for Theoretical Physics, Waterloo, Ontario N2L 2Y5, Canada}
\affiliation{Department of Physics and Astronomy, University of Waterloo, Waterloo, Ontario N2L 3G1, Canada}

\author{Philipp M. Plewa}
\affiliation{Max-Planck-Institut f\"ur extraterrestrische Physik, 85741 Garching, Germany}

\author{Sotiris Chatzopoulos}
\affiliation{Max-Planck-Institut f\"ur extraterrestrische Physik, 85741 Garching, Germany}

\author{Sheperd S. Doeleman}
\affiliation{MIT Haystack Observatory, Westford, Massachusetts 01886, USA}
\affiliation{Harvard-Smithsonian Center for Astrophysics, Cambridge, Massachusetts 02138, USA}

\author{Frank Eisenhauer}
\affiliation{Max-Planck-Institut f\"ur extraterrestrische Physik, 85741 Garching, Germany}

\author{Vincent L. Fish}
\affiliation{MIT Haystack Observatory, Westford, Massachusetts 01886, USA}

\author{Reinhard Genzel}
\affiliation{Max-Planck-Institut f\"ur extraterrestrische Physik, 85741 Garching, Germany}
\affiliation{Physics and Astronomy Departments, University of California, Berkeley, California 94720 USA}

\author{Ortwin Gerhard}
\affiliation{Max-Planck-Institut f\"ur extraterrestrische Physik, 85741 Garching, Germany}

\author{Michael D. Johnson}
\affiliation{Harvard-Smithsonian Center for Astrophysics, Cambridge, Massachusetts 02138, USA}

\begin{abstract}

In general relativity, the angular radius of the shadow of a black hole is primarily determined by its mass-to-distance ratio and depends only weakly on its spin and inclination. If general relativity is violated, however, the shadow size may also depend strongly on parametric deviations from the Kerr metric. Based on a reconstructed image of Sagittarius~A$^\ast$ (Sgr~A$^\ast$) from a simulated one-day observing run of a seven-station Event Horizon Telescope (EHT) array, we employ a Markov chain Monte Carlo algorithm to demonstrate that such an observation can measure the angular radius of the shadow of Sgr~A$^\ast$ with an uncertainty of $\sim1.5~{\rm \mu as}$ (6\%). We show that existing mass and distance measurements can be improved significantly when combined with upcoming EHT measurements of the shadow size and that tight constraints on potential deviations from the Kerr metric can be obtained.

\end{abstract}


\maketitle

%%%%%%%%%%%%%%%%%%%%%%%%%%%%%%%%%%%%%%%%%%
%\section{INTRODUCTION}

The mass $M$ and distance $D$ of Sagittarius~A$^\ast$ (Sgr~A$^\ast$) have been measured using several techniques. References~\cite{Ghez08,Gillessen09,Meyer12} and Refs.~\cite{Do13,Chatzopoulos15} inferred the mass and distance of Sgr~A$^\ast$ from monitoring stars on orbits around Sgr~A$^\ast$ and in the old Galactic nuclear star cluster, respectively. Combining the results of Refs.~\cite{Gillessen09,Chatzopoulos15} yields the measurements $M=(4.23\pm0.14)\times10^6M_\odot$, $D=8.33\pm0.11~{\rm kpc}$~\cite{Chatzopoulos15}. In addition, the distance of Sgr~A$^\ast$ has been obtained from parallax and proper motion measurements of masers throughout the Milky Way by Ref.~\cite{Reid14} finding $D=8.34\pm0.16~{\rm kpc}$. Sgr~A$^\ast$ is also a prime target of the Event Horizon Telescope (EHT), a global very-long baseline interferometer, which has already resolved structures on scales of only $8r_g$ with a three-station array~\cite{Doele08}, where $r_g\equiv GM/c^2$ is the gravitational radius.

According to the general-relativistic no-hair theorem, stationary, electrically neutral black holes in vacuum only depend on their masses $M$ and spins $J$ and are uniquely described by the Kerr metric. Mass and spin are the first two multipole moments of the Kerr metric, and all higher-order moments can be expressed by the relation $M_l + iS_l = M(ia)^l$, where $M_l$ and $S_l$ are the mass and current multipole moments, respectively, and $a\equiv J/M$ is the spin parameter (see, e.g., Ref.~\cite{heu96}). However, general relativity remains practically untested in the strong-field regime found around compact objects, and a final proof of the Kerr nature of black holes is still lacking~\cite{psalLRR}. 

The no-hair theorem can be tested in a model-independent manner using parametrically deformed Kerr-like spacetimes that depend on one or more free parameters in addition to mass and spin. Observations may then be used to measure the deviations. If none are detected, the compact object is consistent with a Kerr black hole. If, however, nonzero deviations are measured, there are two possible interpretations: 1. general relativity holds but the object is some exotica instead of a black hole and 2. the no-hair theorem is falsified. 

Such Kerr-like spacetimes encompass many theories of gravity at once and the underlying action is usually unknown. Here, we employ the Kerr-like metric of Ref.~\cite{J13metric} with the nonzero (dimensionless) deviation parameters $\alpha_{13}$ and $\beta$. This metric can be mapped to known black hole solutions of alternative gravity theories for certain choices of these (and other) parameters and reduces to the Kerr metric if all deviations vanish~\cite{J13metric,Supplementary}. 

\begin{figure}[ht]
\begin{center}
\psfig{figure=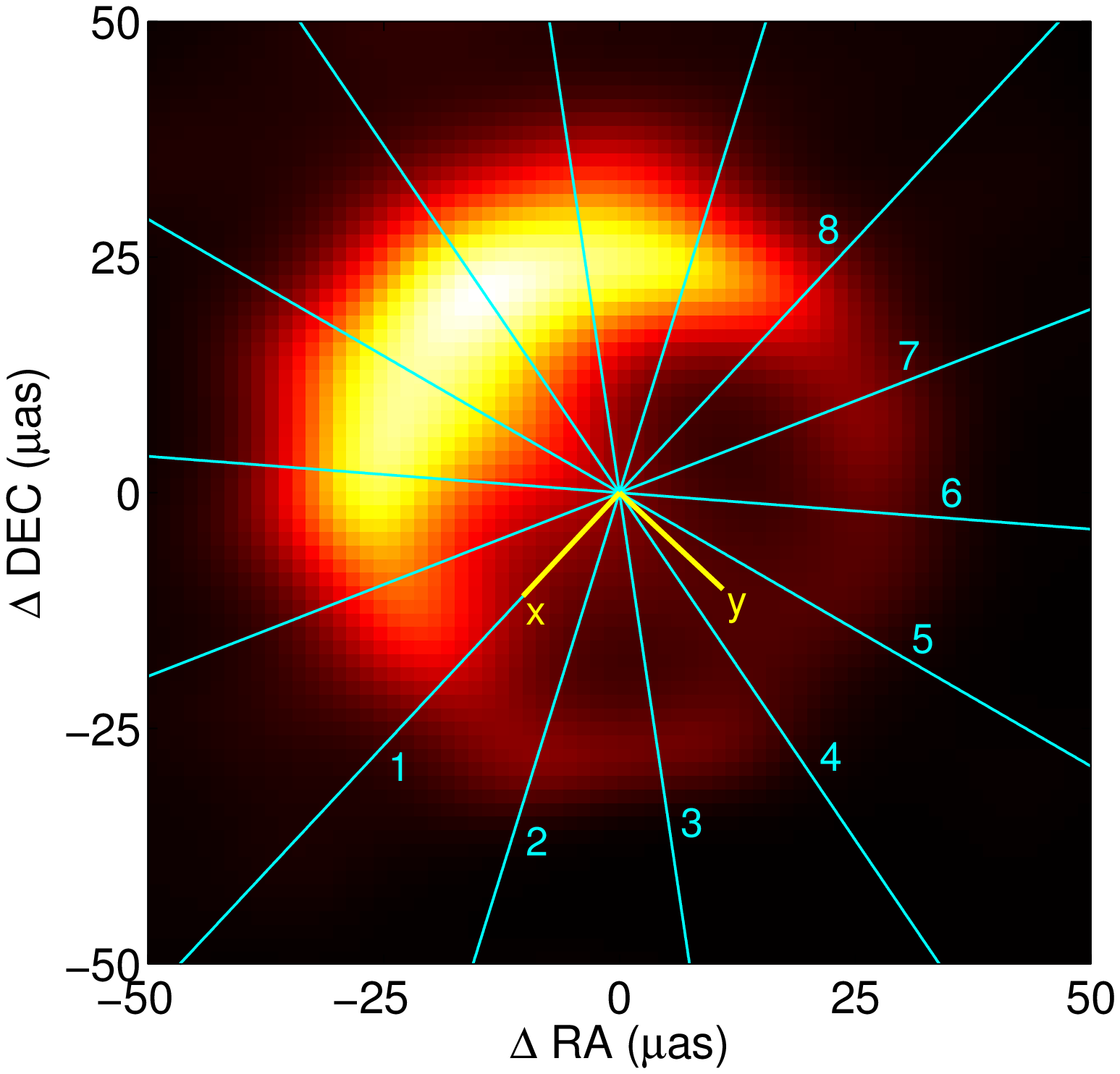,height=2.7in}
\psfig{figure=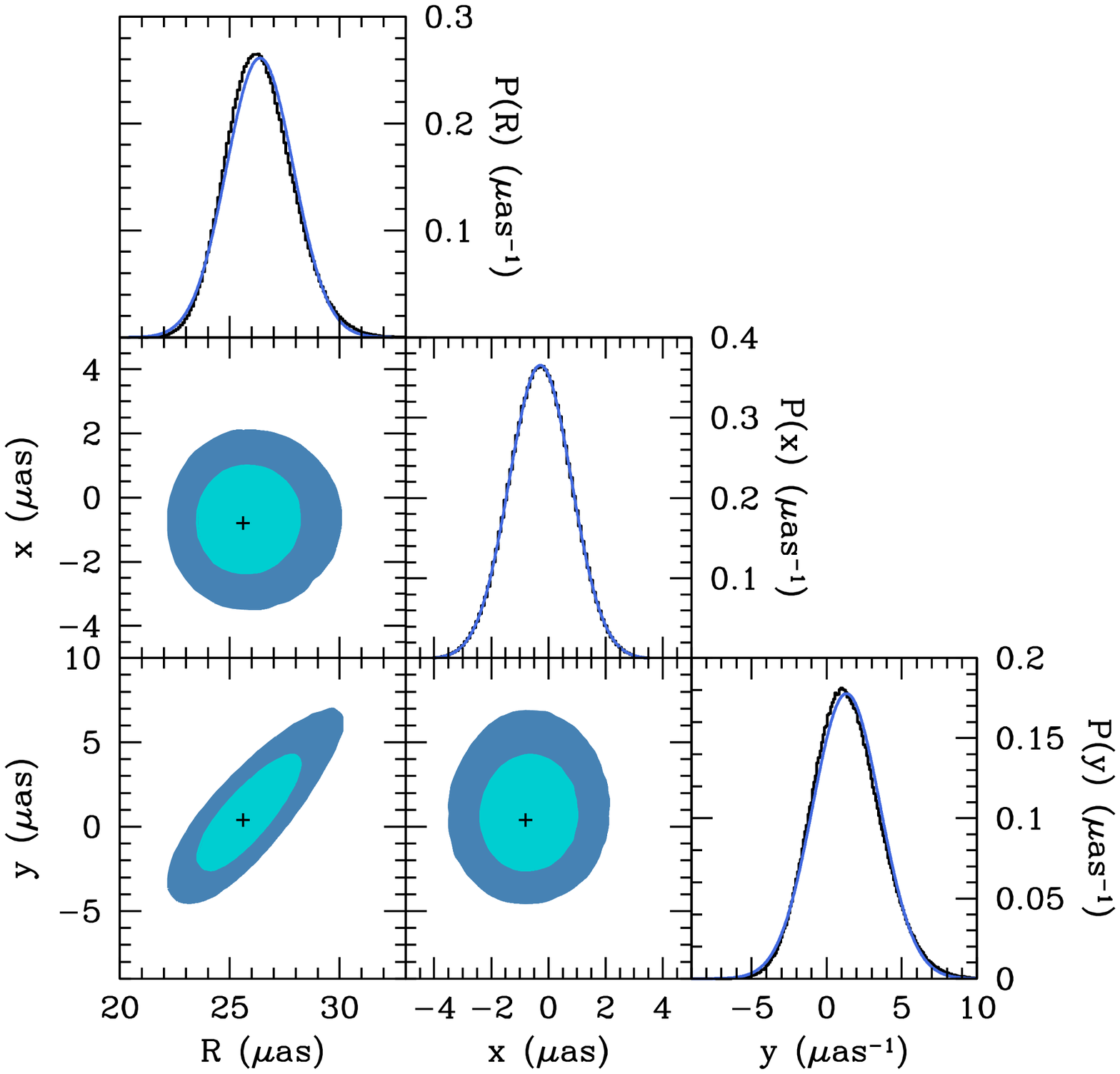,height=3.2in}
\end{center}
\caption{(Top panel) A reconstructed image of Sgr~A$^\ast$ for an EHT observation at 230~GHz with a seven-station array taken from Ref.~\cite{deblurring}. The image shows seven chords for which we determined respective angular radii from Gaussian fits of the intensity profile along the chord sections labeled \mbox{1, $\ldots$, 8}. (Bottom panel) The resulting distributions of the angular radius $R$ of the shadow and the offset $(x,y)$ of the corresponding image center relative to the center of the chords using a Markov chain Monte Carlo sampling of a small region around the center of the chords.}
\label{fig:ringimages}
\end{figure}

A key objective of the EHT is to produce the first direct image of a black hole~\cite{EHT}. These typically reveal a dark region at the center, the so-called shadow, which is surrounded by a bright and narrow ringlike structure embedded within a typically complex image~\cite{shadow}. For a Kerr black hole at a distance $D$, the angular radius $R$ of its shadow is $R\approx5r_g/D$. The shape of this shadow is exactly circular for a Schwarzschild black hole and nearly circular for a Kerr black hole unless its spin is very large and the inclination is high~\cite{PaperII}. However, if the no-hair theorem is violated, the shape of the shadow can become asymmetric~\cite{PaperII} and its size can vary~\cite{AE12,J13rings}.

If Sgr~A$^\ast$ is indeed a Kerr black hole, then its angular radius measured by upcoming EHT observations has to coincide with the angular radius inferred from existing measurements of the mass and distance of Sgr~A$^\ast$~\cite{PsaltisNullTest}. For nonzero values of the parameters $\alpha_{13}$ and $\beta$ in the metric of Ref.~\cite{J13metric}, the shadow size either increases or decreases drastically, retaining a nearly circular shape as long as $a\lesssim0.9r_g$~\cite{AE12,J13rings}, while dynamical measurements of the mass and distance would be unaffected. Thus, the combination of both techniques allows us to constrain the deviation parameters $\alpha_{13}$ and $\beta$. In addition, measurements of the ratio $M/D$ by the EHT can improve upon the existing mass and distance measurements~\cite{SMBHmasses}.

%%%%%%%%%%%%%%%%%%%%%%%%%%%%%%%%%%%%%%
%\section{Radius Estimate}

Here, we investigate the prospects of measuring the angular radius $R$ of the shadow of Sgr~A$^\ast$ with the EHT at 230~GHz assuming a circular shadow. We use a reconstructed image of Sgr~A$^\ast$ based on a simulated image of a radiatively inefficient accretion flow~\cite{deblurring}. The deblurring algorithm of Ref.~\cite{deblurring} corrects the distortions of the simulated visibilities by interstellar scattering so that the resolution of the image is predominantly determined by the instrumental beam.

The top panel of Fig.~\ref{fig:ringimages} shows the reconstructed image of Ref.~\cite{deblurring} using the data of a simulated one-day observing run with a seven-station EHT array assuming realistic measurement conditions but neglecting small-scale variability in the image. The large-scale features within the image are asymmetric due to the relativistic motion of the emitting plasma and opacity. Nevertheless, over at least half of the image the photon ring is clearly visible. To estimate its size we first obtain a set of mean radii $\bar{r}_j$ from Gaussian fits of the specific intensity along the chord sections $j=1,\ldots,8$ shown in Fig.~\ref{fig:ringimages}~\cite{Supplementary}.

The chosen center of the chords most likely does not coincide exactly with the true image center which will be shifted by an offset ${\bf x}=(x,y)$ relative to the center of the chords (see Fig.~\ref{fig:ringimages}). A measured angular radius vector ${\bf\bar{r}_j}=(\bar{r}_j\cos\theta,\bar{r}_j\sin\theta)$ is then related to the angular radius vector of the shadow ${\bf R}$ and the offset ${\bf x}$ by the identity ${\bf\bar{r}_j}={\bf R}-{\bf x}$, i.e., $\bar{r}_j = -(x\cos\theta + y\sin\theta) + \sqrt{ (x\cos\theta + y\sin\theta)^2 + R^2 - (x^2+y^2) }$.

\begin{figure*}[ht]
\begin{center}
\psfig{figure=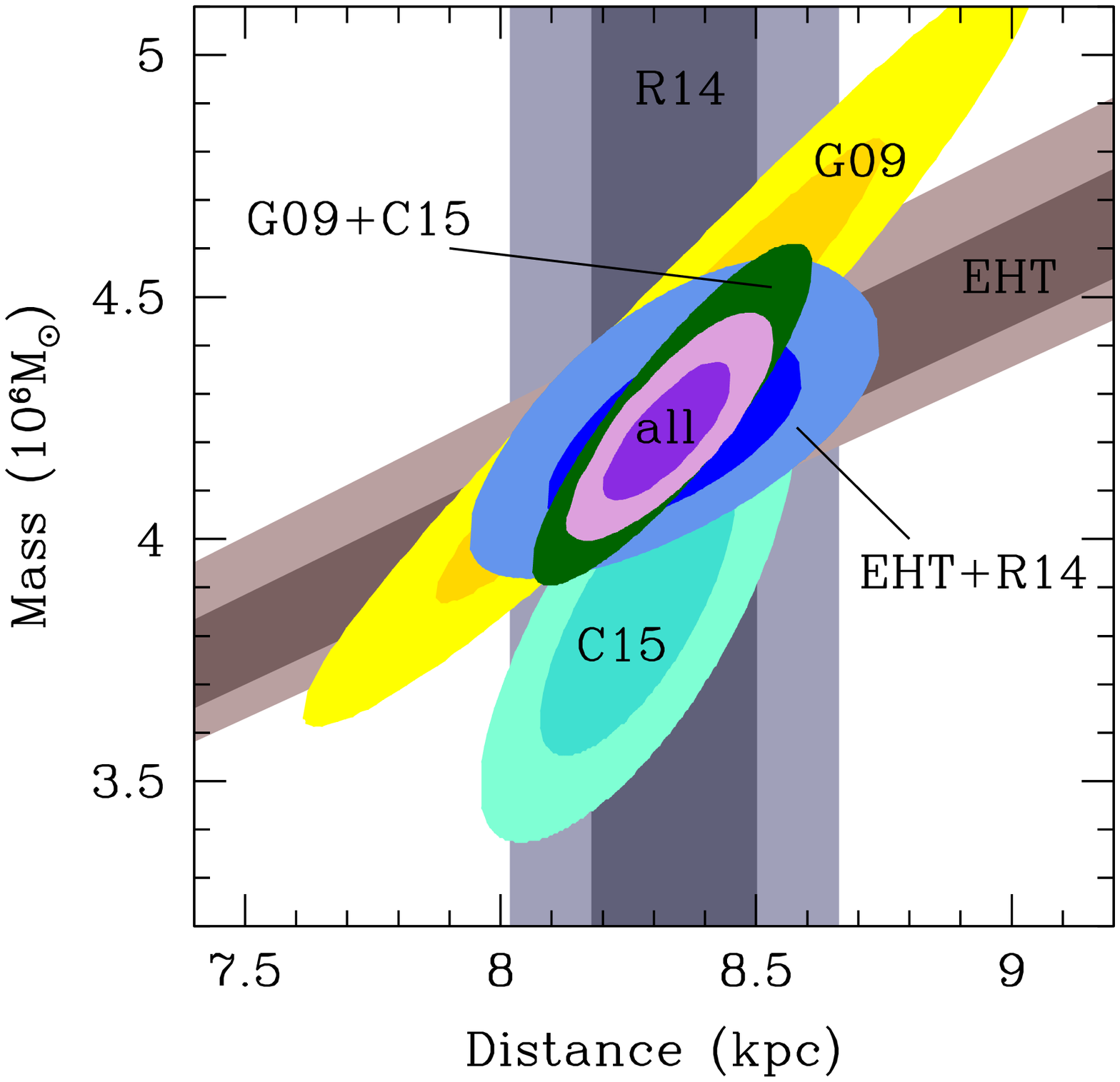,height=2.32in}
\psfig{figure=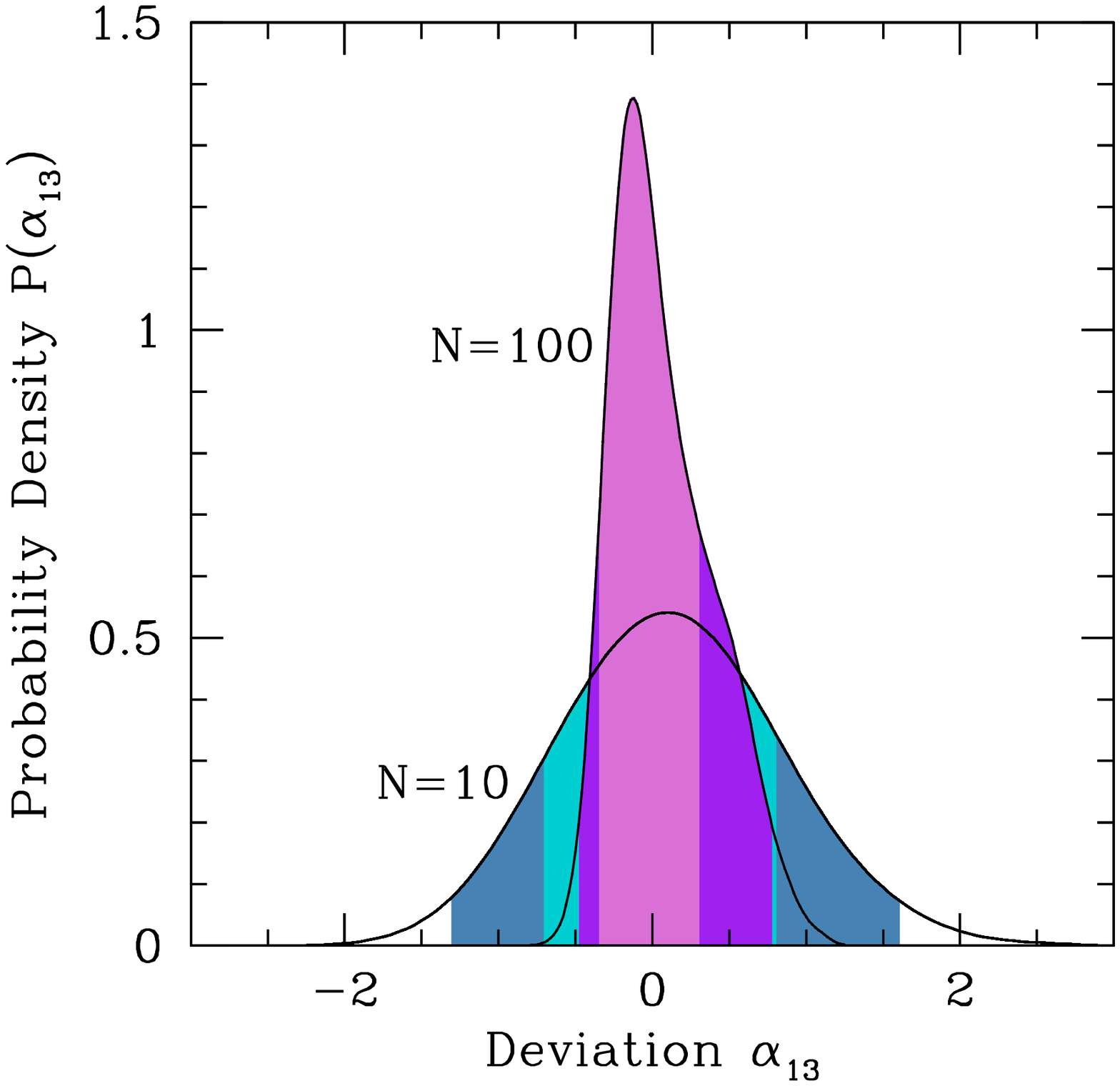,height=2.32in}
\psfig{figure=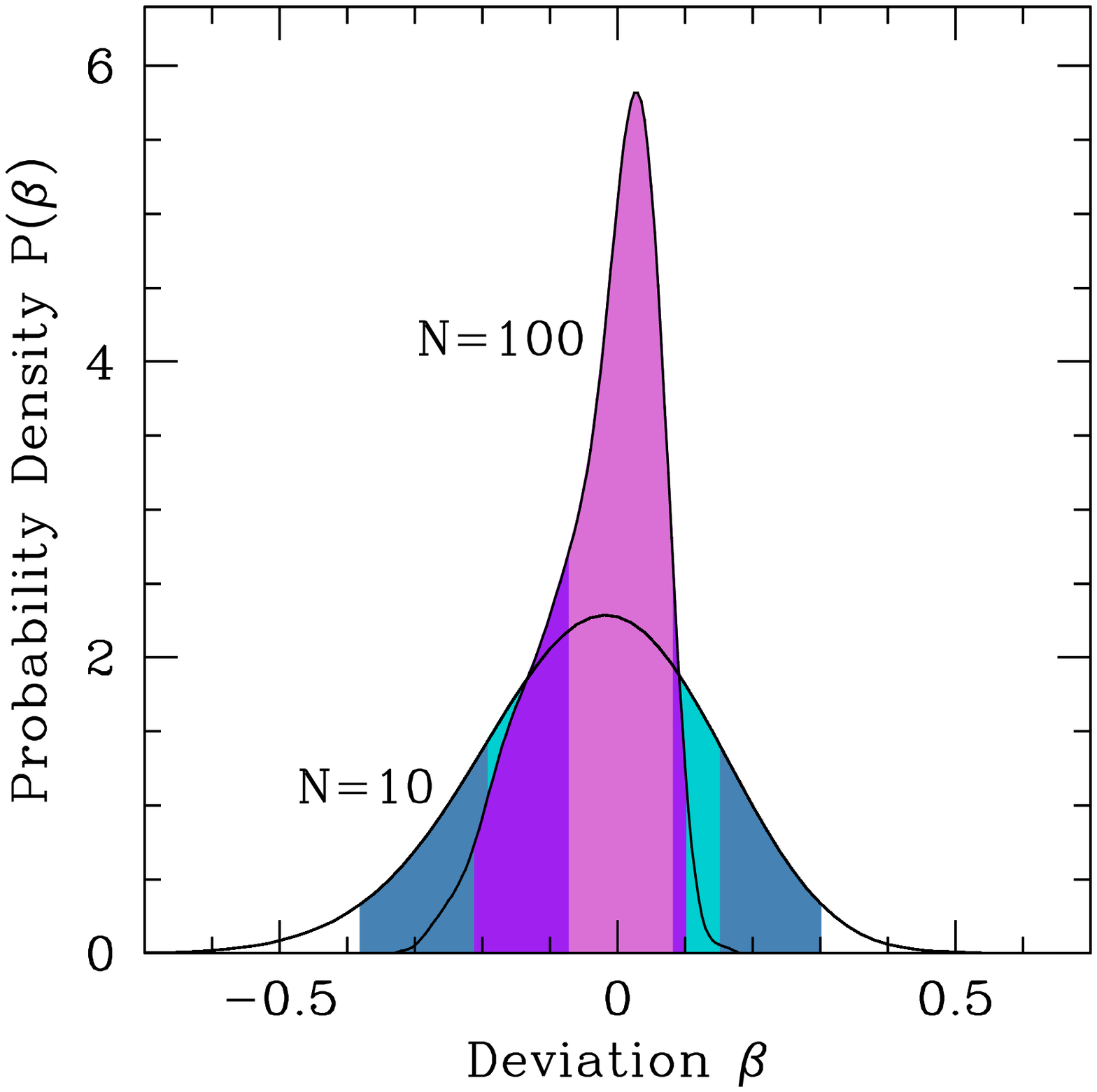,height=2.32in}
\end{center}
\caption{(Left panel) $1\sigma$ and $2\sigma$ confidence contours of the probability density of the mass and distance of Sgr~A$^\ast$ for existing measurements ({\it S}~stars, ``G09''~\cite{Gillessen09}; masers, ``R14''~\cite{Reid14}; star cluster, ``C15''~\cite{Chatzopoulos15}), a simulated measurement of the shadow size of Sgr~A$^\ast$ for $N=10$ observations with a seven-station EHT array (``EHT''), and several combinations thereof. The simulated EHT measurement improves the other constraints on the mass and distance significantly. (Center and right panels) Simulated $1\sigma$ and $2\sigma$ confidence contours of the probability density of the deviation parameters $\alpha_{13}$ and $\beta$, respectively, corresponding to $N=10$ and $N=100$ EHT observations, each marginalized over the mass and distance using the combination of all data sets (``all'') in the $N=10$ case and of simulated stellar-orbit observations from a 30-m-class telescope~\cite{Weinberg95} in the $N=100$ case.}
\label{fig:constraints}
\end{figure*}

From the set of angular radii $\bar{r}_j$ we then obtain estimates of the angular shadow radius $R$ and the offsets $x$ and $y$ using a Markov chain Monte Carlo sampling of a small region around the center of the chords. We find that each of the resulting distributions of the angular shadow radius $R$ and the offsets $x$ and $y$ is approximately Gaussian and that the angular radius $R$ is correlated with the offset $y$, while both the angular radius $R$ and the offset $y$ appear to be uncorrelated with the offset $x$ (see Fig.~\ref{fig:constraints}). Thus, we infer $x=(-0.3\pm1.1)~{\rm \mu as}$, $y=(1.3\pm2.2)~{\rm \mu as}$, and $R=(26.4\pm1.5)~{\rm \mu as}$, which corresponds to a distance $\sim0.16r_g$ and a $\sim6\%$ uncertainty of the angular radius. This is comparable to the one of Ref.~\cite{PsaltisNullTest}, whose authors used a pattern matching technique. 

Although we find no significant values of the offsets $x$ and $y$, we repeated this process several times by slightly shifting the center of the chords and obtained a similar result in each case. Our estimate of the angular radius is consistent with the actual angular radius of the shadow $R\approx27.6~{\rm \mu as}$ at the $1\sigma$ level corresponding to the values of the mass $M=4.3\times10^6M_\odot$, distance $D=8~{\rm kpc}$, and spin $a=0$ used in the simulated image shown in Fig.~\ref{fig:ringimages}. Because the radius estimate would be exact for a true image~\cite{pathlength}, our method seems to have no significant bias.

%%%%%%%%%%%%%%%%%%%%%%%%%%%%%%%%%%%%%%
%\section{Constraints}

We combine the above simulated EHT measurement of the angular shadow radius of Sgr~A$^\ast$ with existing measurements of its mass and distance which we use as a prior $P_{\rm prior}(M,D)$. We assume a Gaussian distribution $P_{\rm EHT}({\rm data}|M,D,a,\vartheta,\alpha_{13},\beta)$ of the angular radius with an uncertainty $\sigma=1.5~{\rm \mu as}$ and a mean corresponding to the maximum of the distribution $P_{\rm prior}(M,D)$ of the combined measurements of Refs.~\cite{Gillessen09,Chatzopoulos15,Reid14} assuming a Kerr black hole with spin $a=0.5r_g$ and inclination $\vartheta=60^\circ$. For given values of the mass, distance, and deviation parameters, we calculate the angular radius of the shadow as described in Refs.~\cite{AE12,J13rings}. Because the shadow size depends only weakly on the spin and inclination, we marginalize over the spin and inclination using Eq.~(42) of Ref.~\cite{J13rings} in the case of a Kerr black hole and a fine grid of points $(a,\vartheta)$ in the case of non-Kerr black holes. Last, we use Bayes's theorem to express the likelihood of the mass, distance, and deviation parameters given the data as $P(M,D,\alpha_{13},\beta|{\rm data}) = C~P_{\rm EHT}({\rm data}|M,D,\alpha_{13},\beta)P_{\rm prior}(M,D)$, where $C$ is a normalization constant. For simplicity, we also consider the modifications of the shadow size introduced by the parameters $\alpha_{13}$ and $\beta$ separately.

Joint constraints from various measurements of $M$ and $D$ are combined in Fig.~\ref{fig:constraints}; where indicated, we presume forthcoming 30-m-class telescope stellar dynamics observations will make roughly 0.1\% measurements of $M$ and $D$~\cite{Weinberg95}. Here, we assume that all EHT measurements are independent and identical allowing us to reduce their uncertainty by a factor of $\sqrt{N}$. In practice, one EHT measurement likely corresponds to an average of several observations due to small-scale variations in the image which we neglect here.

We find that the EHT alone can measure the mass-to-distance ratio (in units of $10^6M_\odot/{\rm kpc}$) $M/R=0.505^{+0.013+0.029}_{-0.011-0.020}$ for $N=10$ observations and $M/R=0.502^{+0.010+0.026}_{-0.005-0.007}$ for $N=100$ observations, respectively. Table~\ref{tab:I} lists constraints on the mass and distance corresponding to various combinations of the EHT measurements for ten observations with existing data showing significant improvements, facilitated by different correlations of mass and distance (stellar orbits: roughly $M\sim D^2$~\cite{Ghez08,Gillessen09}; EHT: $M\sim D$~\cite{SMBHmasses}). In particular, combining the EHT result with the parallax measurement by Ref.~\cite{Reid14} is comparable to the mass and distance measurements from stellar orbits including the combined result of Refs.~\cite{Gillessen09,Chatzopoulos15}. Combining all data sets, we obtain the constraints on the deviation parameters $\alpha_{13}=0.1^{+0.7+1.5}_{-0.8-1.4}$, $\beta=-0.02^{+0.17+0.32}_{-0.17-0.36}$ in the $N=10$ case, while, in the $N=100$ case, we find $\alpha_{13}=-0.13^{+0.43+0.90}_{-0.21-0.34}$, $\beta=0.03^{+0.05+0.07}_{-0.10-0.24}$; the uncertainties of the mass and distance remain at the $\sim0.1\%$ level. Here, all results are quoted with $1\sigma$ and $2\sigma$ error bars, respectively. Our results would also imply constraints on the couplings of specific gravity theories, summarized in Table~\ref{tab:III}, and could improve upon spin measurements by other methods~\cite{Supplementary}.

\begin{table}[ht]
\begin{center}
\footnotesize
\begin{tabular}{lcc}
\multicolumn{3}{c}{}\\
Data   & ~~Mass ($10^6M_\odot$) & ~~Distance (kpc) \\
\hline \hline
EHT+G09 & ~~$4.16^{+0.18+0.38}_{-0.16-0.31}$ & ~~$8.18^{+0.19+0.39}_{-0.19-0.37}$ \\
EHT+R14 & ~~$4.22^{+0.13+0.28}_{-0.13-0.24}$ & ~~$8.34^{+0.16+0.32}_{-0.15-0.31}$ \\
EHT+C15 & ~~$4.17^{+0.11+0.22}_{-0.11-0.21}$ & ~~$8.38^{+0.11+0.21}_{-0.11-0.21}$ \\
All     & ~~$4.22^{+0.09+0.20}_{-0.09-0.17}$ & ~~$8.33^{+0.08+0.17}_{-0.08-0.15}$ \\
\hline
\end{tabular}
\caption{Simulated mass and distance measurements using existing data (G09~\cite{Gillessen09}, R14~\cite{Reid14}, C15~\cite{Chatzopoulos15}) as priors.}
\label{tab:I}
\end{center}
\end{table}

\begin{table}[h]
\begin{center}
\footnotesize
\begin{tabular}{lll}
\multicolumn{3}{c}{}\\
Theory  & ~Constraints ($N=10$) & ~Constraints ($N=100$)  \\
\hline \hline
RS2	&  ~$\beta_{\rm tidal}=-0.02^{+0.17+0.32}_{-0.17-0.36}$  & ~$\beta_{\rm tidal}=0.03^{+0.05+0.07}_{-0.10-0.24}$  \\
MOG     &  ~$\alpha=-0.02^{+0.13+0.22}_{-0.13-0.24}$             & ~$\alpha=0.03^{+0.05+0.06}_{-0.08-0.17}$  \\
EdGB    &  ~$\zeta_{\rm EdGB}\approx0^{+0.1+0.2}_{-0.1-0.3}$     & ~$\zeta_{\rm EdGB}\approx0.022^{+0.035+0.057}_{-0.072-0.150}$ \\
\hline
\end{tabular}
\caption{
%$1\sigma$ and $2\sigma$ constraints on the parameters of black holes in specific theories of modified gravity implied by our simulation.
$1\sigma$ and $2\sigma$ constraints on the parameters of black holes in Randall-Sundrum-type braneworld gravity (RS2~\cite{RS2BH}), modified gravity (MOG~\cite{MOG}), and Einstein-dilaton-Gauss-Bonnet gravity (EdGB~\cite{EdGB}) implied by our simulation.
}
\label{tab:III}
\end{center}
\end{table}

%%%%%%%%%%%%%%%%%%%%%%%%%%%%%%%%%%%%%%
%\section{Discussion}

We have assumed that the image of Sgr A$^\ast$ is constant, but two effects will cause the apparent structure of the source to be variable. First, the accretion flow will be intrinsically variable on characteristic time scales on the order of minutes to tens of minutes. These rapid changes will complicate image reconstructions from EHT data, which conventionally assume a static source over each observing epoch. Second, interstellar scattering will blur the source image, which reduces the sensitivity of the long baselines of the EHT but is largely invertible. Since we fit the specific intensity along the chords with Gaussians, our method is insensitive to remaining uncertainties in the interstellar scattering law. Interstellar scattering will also introduce refractive substructure into the apparent image which can cause image distortions that will vary stochastically from epoch to epoch~\cite{deblurring,Johnson15}.

While intrinsic source variability and refractive substructure are very different processes, their primary mitigation strategy is to repeat observations over many epochs. It will be possible to image the quiescent accretion flow by imaging data obtained over the course of several nights, over which time the source variability will average out~\cite{Lu15}. Likewise, the effects of different realizations of refractive substructure on different observing days will average out.

Our estimate will also be affected moderately by systematic uncertainties in the EHT calibration and the accretion flow model in the image reconstruction. In order to properly account for the calibration uncertainties in our estimate, perhaps the best approach is to reconstruct images using not only the best-calibrated data but also a number of data sets that are intentionally miscalibrated by a small amount. While the detailed structure of the accretion flow of Sgr~A$^\ast$ is unknown and could have, e.g., other (vertical or horizontal) configurations of the plasma density and magnetic fields, as well as outflows, we expect that these uncertainties only play a subdominant role in our simulation as long as a (nearly circular) shadow is clearly visible in the image, because the size and shape of the shadow are almost entirely determined by the underlying spacetime alone~\cite{Supplementary}.

Before a 30-m-class optical telescope will be available, the uncertainties of mass and distance measurements based on stellar orbits will be further reduced by continued monitoring and the expected improvement in astrometry with the second-generation instrument GRAVITY for the Very Large Telescope Interferometer~\cite{GRAVITY}.

%%%%%%%%%%%%%
% acknowledgments

T. J. is supported in part by Perimeter Institute for Theoretical Physics. A. E. B. receives financial support from Perimeter Institute for Theoretical Physics and the Natural Sciences and Engineering Research Council of Canada through a Discovery Grant. Research at Perimeter Institute is supported by the Government of Canada through Industry Canada and by the Province of Ontario through the Ministry of Research and Innovation. The Event Horizon Telescope is supported by grants from the National Science Foundation, from the Gordon and Betty Moore Foundation (Grant No. GBMF-3561), from the Smithsonian Institution, and with generous equipment donations from Xilinx Inc. and HGST Inc.

%%%%%%%%%%%%%%%%%%%%%%%%%%%%%%%%%%%%%%%

%%% Supplementary Material
\vspace{0.4cm}

\begin{center}
{\large Supplemental Material}
\end{center}
\vspace{0.2cm}

{\it Generality of Kerr-like metrics ---} Several Kerr-like metrics have been proposed to date (e.g.,~\cite{GB06,VH10,VYS11,JPmetric,J13metric}). Among these, the metric of Ref.~\cite{J13metric} is the only non-perturbative metric (i.e., it is not limited to the description of only small perturbations away from the Kerr metric) that depends on (at least) four free, independent functions that parameterize potential deviations from the Kerr metric. Since in general relativity stationary, axisymmetric, asymptotically flat, vacuum metrics generally depend on only four functions~\cite{Wald84}, Kerr-like metrics that possess these properties should depend on at least four such functions. Moreover, this metric harbors a black hole, whereas the metrics of Refs.~\cite{GB06,VH10,JPmetric} generally harbor naked singularities as well as, in the case of the metrics of Refs.~\cite{GB06,VH10}, unphysical regions where causality is violated~\cite{pathologies}. This motivates our choice of the metric of Ref.~\cite{J13metric} for our analysis.

Kerr-like metrics generally do not derive from the action of any particular gravity theory. Instead, they are a class of so-called metric theories of gravity~\cite{MTW} and typically obey the full Einstein equivalence principle (EEP)~\cite{WillLRR}. Insight into the underlying theory of gravity is then hoped to be gained through observations. The EEP is the foundation of such a theory and is comprised of three fundamental principles, the weak equivalence principle (WEP), local Lorentz invariance (LLI), and local position invariance (LPI). The WEP postulates that the trajectory of a freely falling ``test'' body, i.e., a body that is not affected by forces such as electromagnetism or tidal gravitational forces, is independent of its internal structure and composition. In Newtonian gravity, this statement is equivalent to the equality of the inertial and gravitational mass of such a test body. The LLI states that the outcome of any local non-gravitational experiment is independent of the velocity of the freely-falling reference frame in which it is performed. The LPI postulates that the outcome of such an experiment is independent of its position and the time of its performance~\cite{WillLRR}. It then follows from the EEP that gravitation can be described by the curvature of a spacetime (e.g.,~\cite{Will93}).

The only theories of gravity that are consistent with the EEP are metric theories of gravity~\cite{WillLRR}. In these metric theories, the spacetime is endowed with a symmetric metric, the trajectories of freely falling test bodies are geodesics of that metric, and in local freely falling reference frames, the non-gravitational laws of physics are those of special relativity~\cite{MTW}. This setup, then, allows for the calculation and prediction of possible observable signatures of the theory such as the shape and size of the shadow of a black hole. The three components of the EEP have been thoroughly tested by many different experiments, at least in the weak-field regime~\cite{WillLRR}. Some requirements of the EEP, however, can be relaxed, such as the LLI for black holes in Lorentz-violating theories~\cite{YYT12}.

{\it Implications for alternative gravity theories ---} Our simulated constraints on the deviation parameters $\alpha_{13}$ and $\beta$ translate into specific constraints on the parameters of known black-hole metrics in other theories of gravity. In Randall-Sundrum type braneworld gravity (RS2), black holes also carry a (dimensionless) tidal charge $\beta_{\rm tidal}$~\cite{RS2BH} which can be mapped trivially to the parameter $\beta$ via the equation $\beta=\beta_{\rm tidal}$~\cite{J13metric}. In Modified Gravity (MOG), the parameter $\beta$ relates to the coupling constant $\alpha$ via the equation $\beta=\alpha(1 + \alpha)^2$~\cite{MOG}. The parameter $\alpha_{13}$ can be mapped to the coupling constant $\zeta_{\rm EdGB}$ of slowly rotating black holes [i.e., up to $\mathcal{O}(a)$] in Einstein-dilaton-Gauss-Bonnet gravity (EdGB;~\cite{EdGB}) via the equation $\alpha_{13} \approx -\zeta_{\rm EdGB}/6$. This mapping is only approximate, because the higher-order parameters $\alpha_{1i}$, $i\geq4$, as well as the parameters $\alpha_{2i}$, $i\geq2$, would likewise have to be constrained in order for the mapping to be exact~\cite{Jreview}. We summarize the constraints on the above parameters from our results for the parameters $\alpha_{13}$ and $\beta$ in Table~\ref{tab:III}.

Constraints on potential deviations from the Kerr metric may also improve upon measurements of the spin of Sgr~A$^\ast$ by other observations for several reasons. If no deviations are detected, the Kerr metric itself is validated which is usually an underlying assumption of such spin measurements. However, if a deviation is detected, it has to be incorporated into the spin measurement, at least for spin measurements that are based on observations of the accretion flow. In these measurements, the spin and the deviation parameters are typically strongly correlated (see, e.g., Ref.~\cite{Bro14}). In our case, however, the effect of the spin is minor, so that EHT measurements of the size of the shadow can substantially reduce that correlation. Alternatively, within general relativity, the deviation parameters can also be interpreted as a measure of the astrophysical uncertainties of the accretion flow so that their effects can be treated in a quantitative manner.

Although we have only discussed a prior on the mass and distance of Sgr~A$^\ast$ based on the monitoring of stellar orbits in our analysis, other potential measurements could be used as priors on the mass and distance, as well as on the spin, inclination, and deviation parameters. These include gravitational-wave observations of extreme mass-ratio inspirals into Sgr~A$^\ast$~\cite{GairLRR}, electromagnetic observations of variability (e.g.,~\cite{QPOs}), timing observations of pulsars orbiting around Sgr~A$^\ast$~\cite{Liu12}, and x-ray observations of spectra and iron lines~(e.g.,~\cite{Xrayprobes}).

{\it Summary of and relationship to other EHT gravity tests ---} The results presented here are part of an ongoing effort to test general relativity with observations of Sgr~A$^\ast$ using the EHT. Ref.~\cite{PsaltisNullTest} pursued a similar idea, but there are several important differences between our and their analyses. First, our method is perhaps easier to understand than the more involved method of Ref.~\cite{PsaltisNullTest} who used a pattern matching technique based on a Hough/Radon transform in order to determine the angular radius of the shadow. We simply infer this radius from the peak fluxes along different chords across the image. Second, Ref.~\cite{PsaltisNullTest} infer the angular radius of the shadow from an idealized theoretical image of an accretion flow surrounding Sgr~A$^\ast$ which does not take into account the blurring of the image due to the scattering of photons off free electrons along the line of sight as well as the reconstruction of the image based on a realistic EHT observing run. These effects, however, are critical for a reliable estimate of a measurement of the shadow radius. These uncertainties are discussed in some detail in Ref.~\cite{PsaltisNullTest} and will most likely have a noticeable effect on their radius estimate. In contrast, as mentioned earlier, our estimate is insensitive to these uncertainties. Third, Ref.~\cite{PsaltisNullTest} only discussed the possibility of performing a null test of general relativity within which the mass-to-distance ratio of Sgr~A$^\ast$ inferred from the angular radius of its shadow has to coincide with the one inferred from the monitoring of stellar orbits. Our analysis goes significantly beyond such a null test by quantifying potential deviations from the Kerr metric within the context of a Kerr-like background. Fourth, the radius of the shadow also depends on the spin and the inclination of Sgr~A$^\ast$. Although the angular radius of the shadow is primarily determined by the mass-to-distance ratio, the effect of the spin and the inclination has to be taken into account in an effort to make a realistic prediction of a test of general relativity based on the size of the shadow.

Likewise, Ref.~\cite{Bro14} constrained a potential deviation from the Kerr metric within the context of a specific radiatively-inefficient accretion flow (RIAF) model using early (2007--2009) EHT data~\cite{Doele08,Fish11}. Their constraints on such a deviation were weak and the corresponding confidence contours covered practically the entire parameter space they considered. Our simulated constraints are much stronger and have an unprecedented level of precision. In addition, Ref.~\cite{Bro14} employed a quasi-Kerr metric~\cite{GB06} which only depends on one additional parameter whereas our analysis is based on the metric of Ref.~\cite{J13metric} which depends on (at least) five additional parameters and is much more general in that sense. Only three of these parameters ($\alpha_{13}$, $\alpha_{22}$, and $\beta$) affect the size of the shadow. Since the parameter $\alpha_{22}$ alters the size of the shadow only marginally~\cite{J13rings}, we focus on the other two parameters. More importantly, however, our constraints are largely model-independent as we discuss below allowing for a clean test of general relativity without many of the astrophysical uncertainties regarding the structure of the accretion flow.

\begin{figure*}
\begin{center}
\psfig{figure=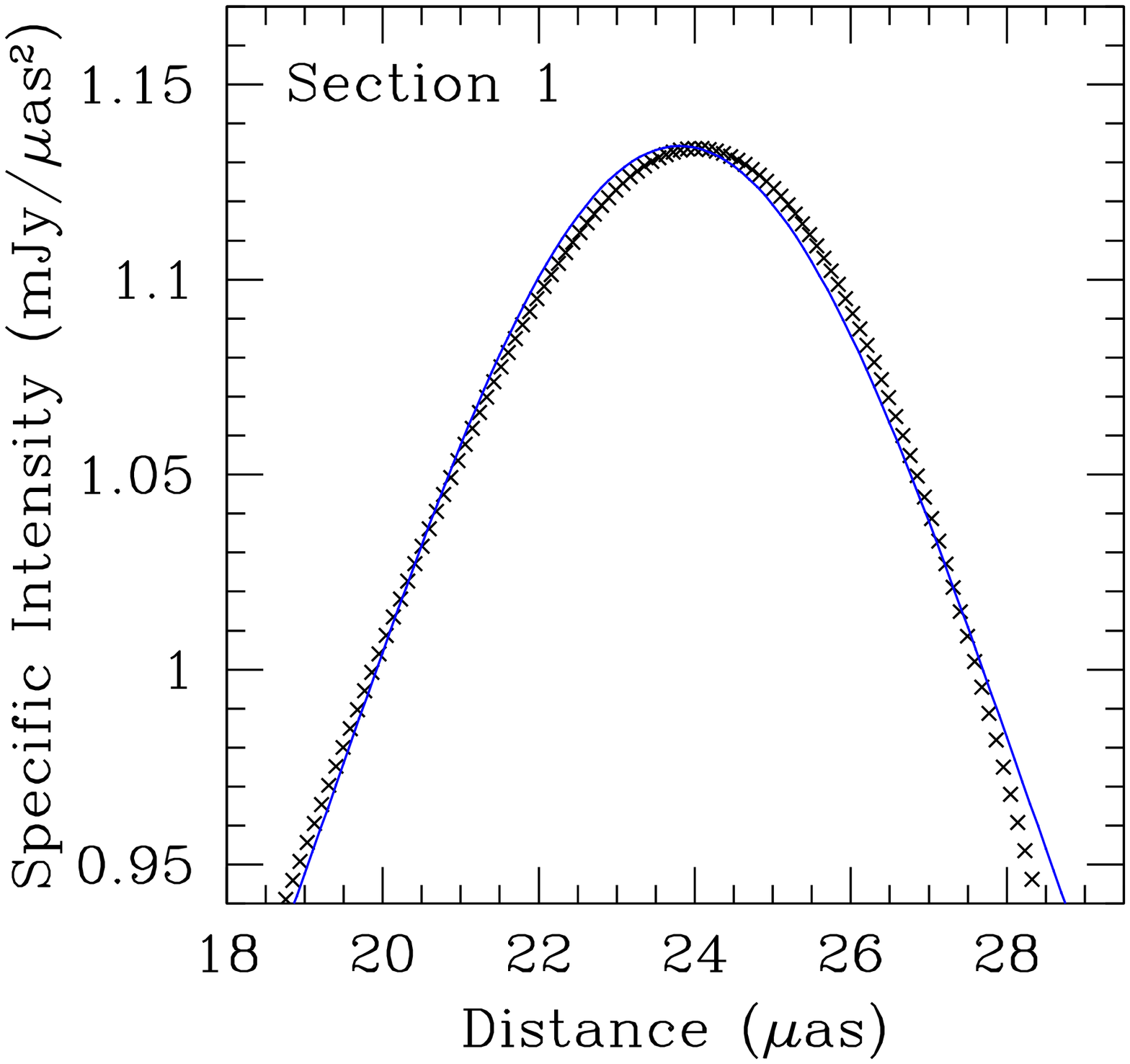,height=1.73in}
\psfig{figure=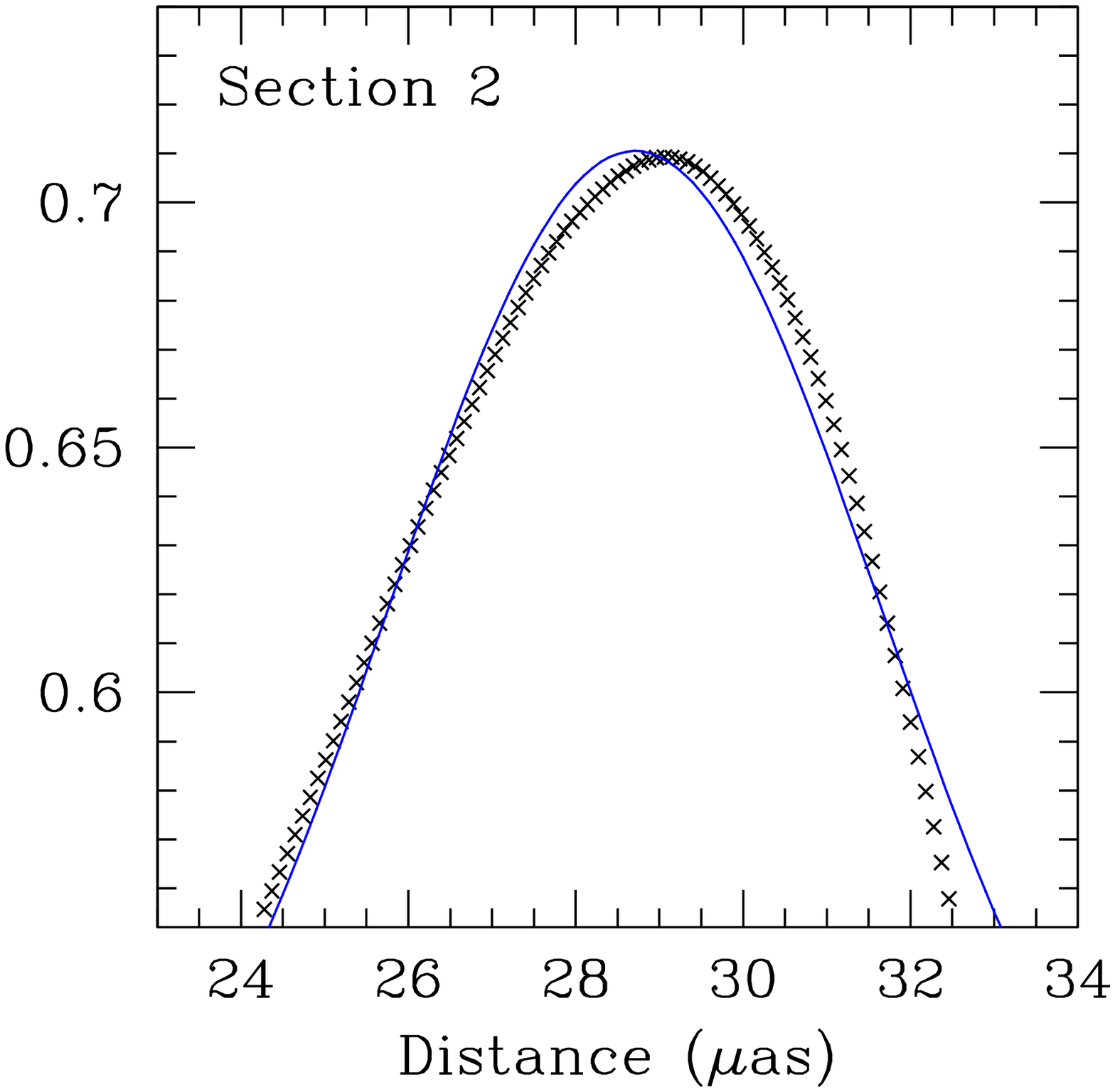,height=1.73in}
\psfig{figure=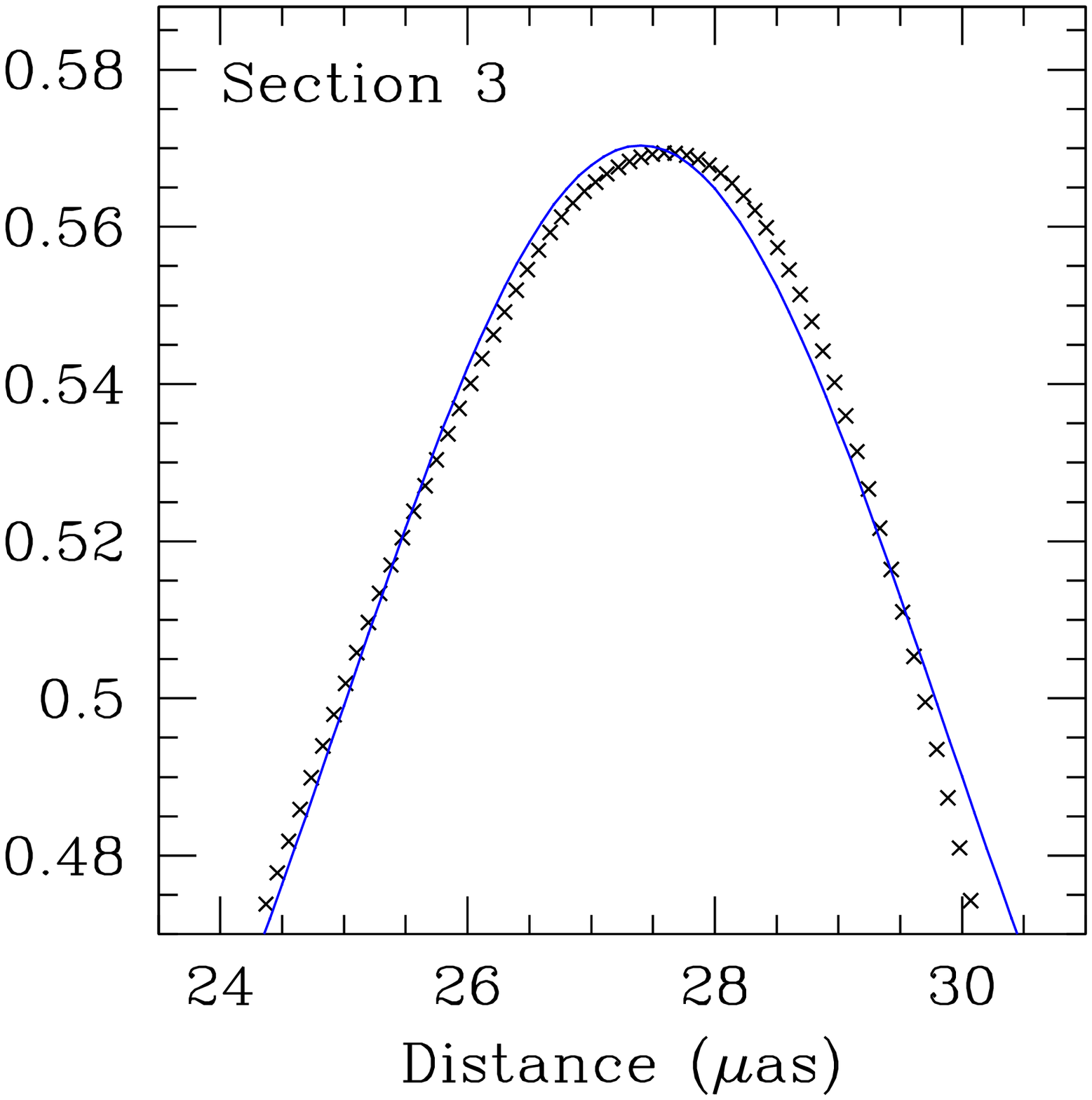,height=1.73in}
\psfig{figure=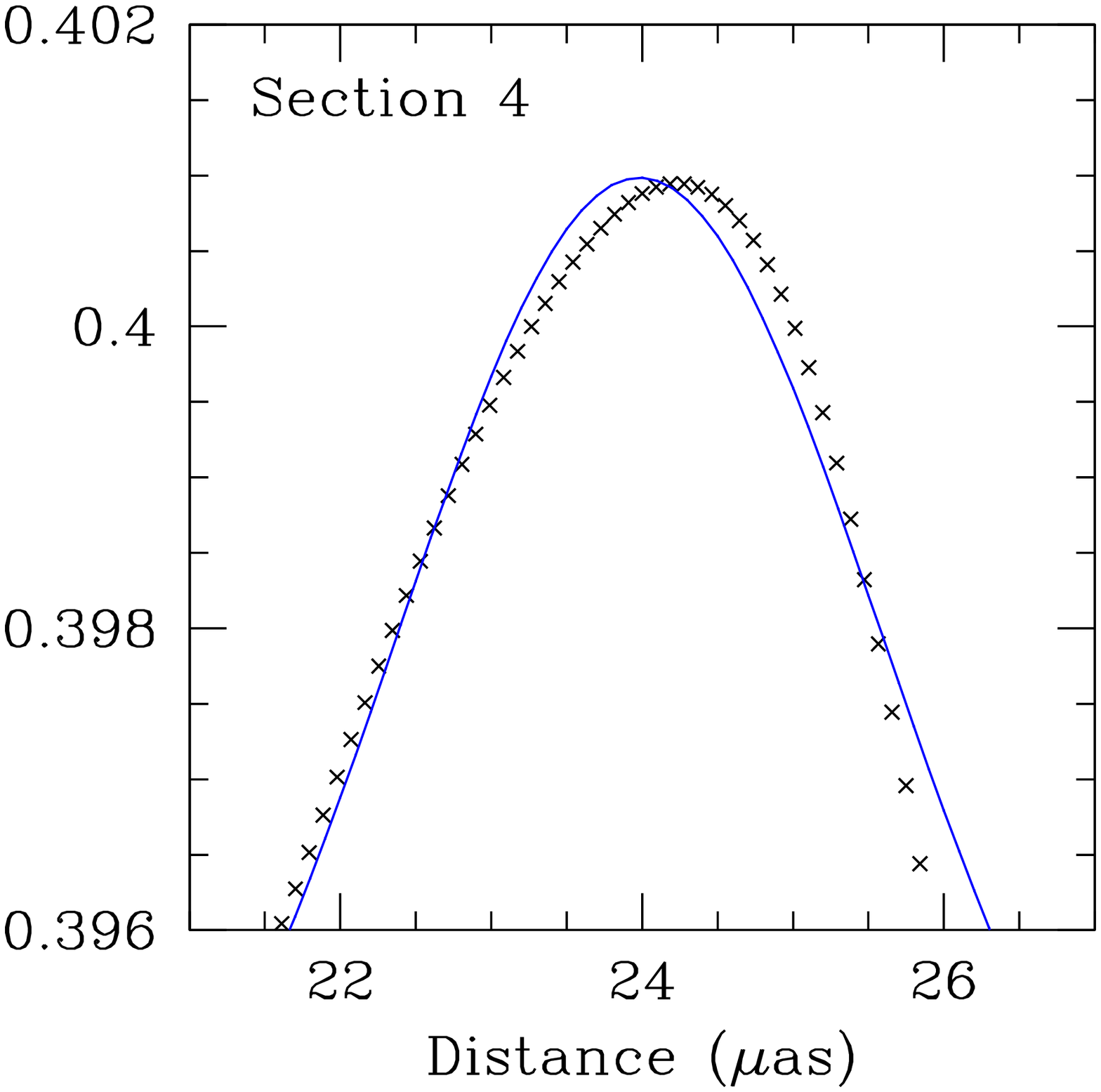,height=1.73in}
\psfig{figure=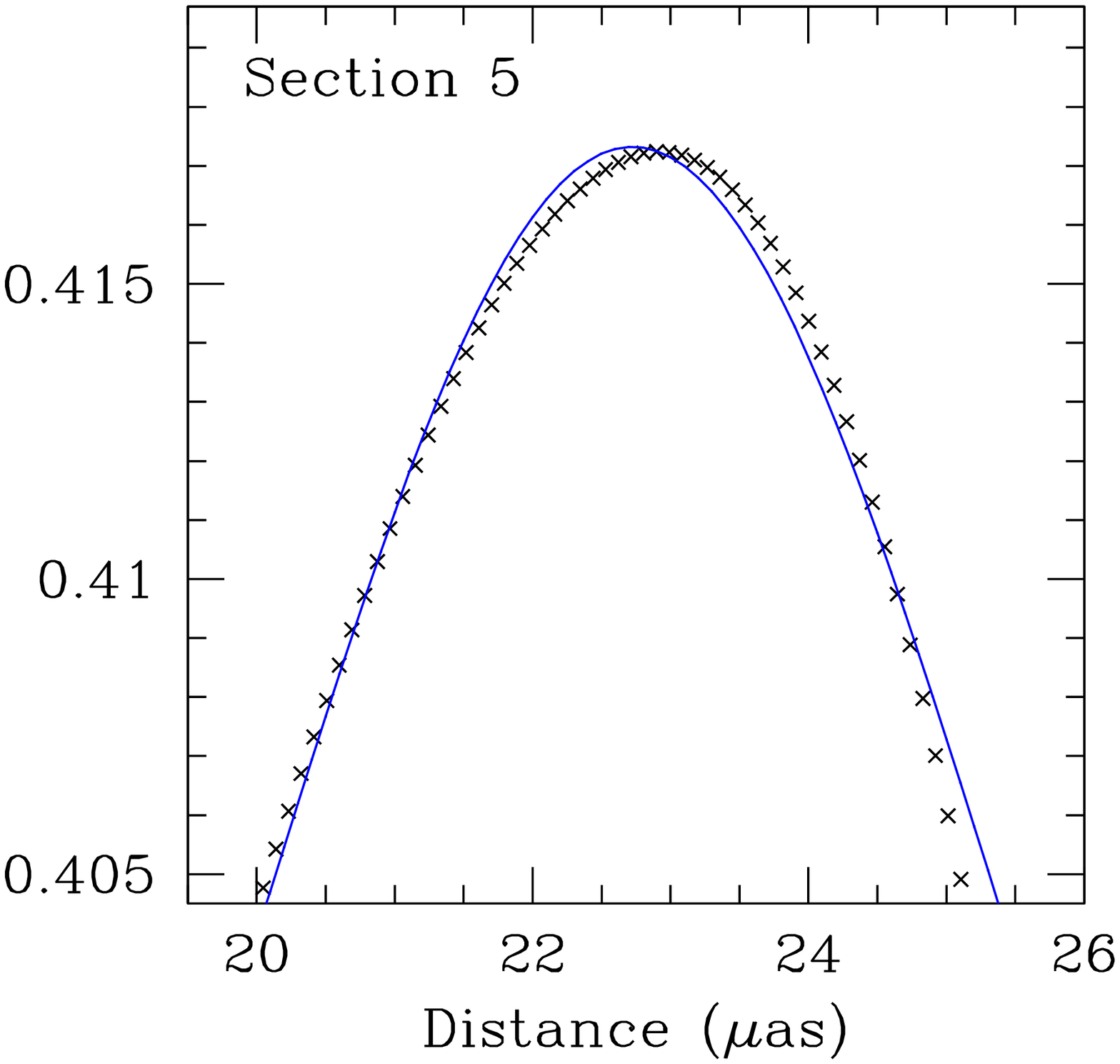,height=1.73in}
\psfig{figure=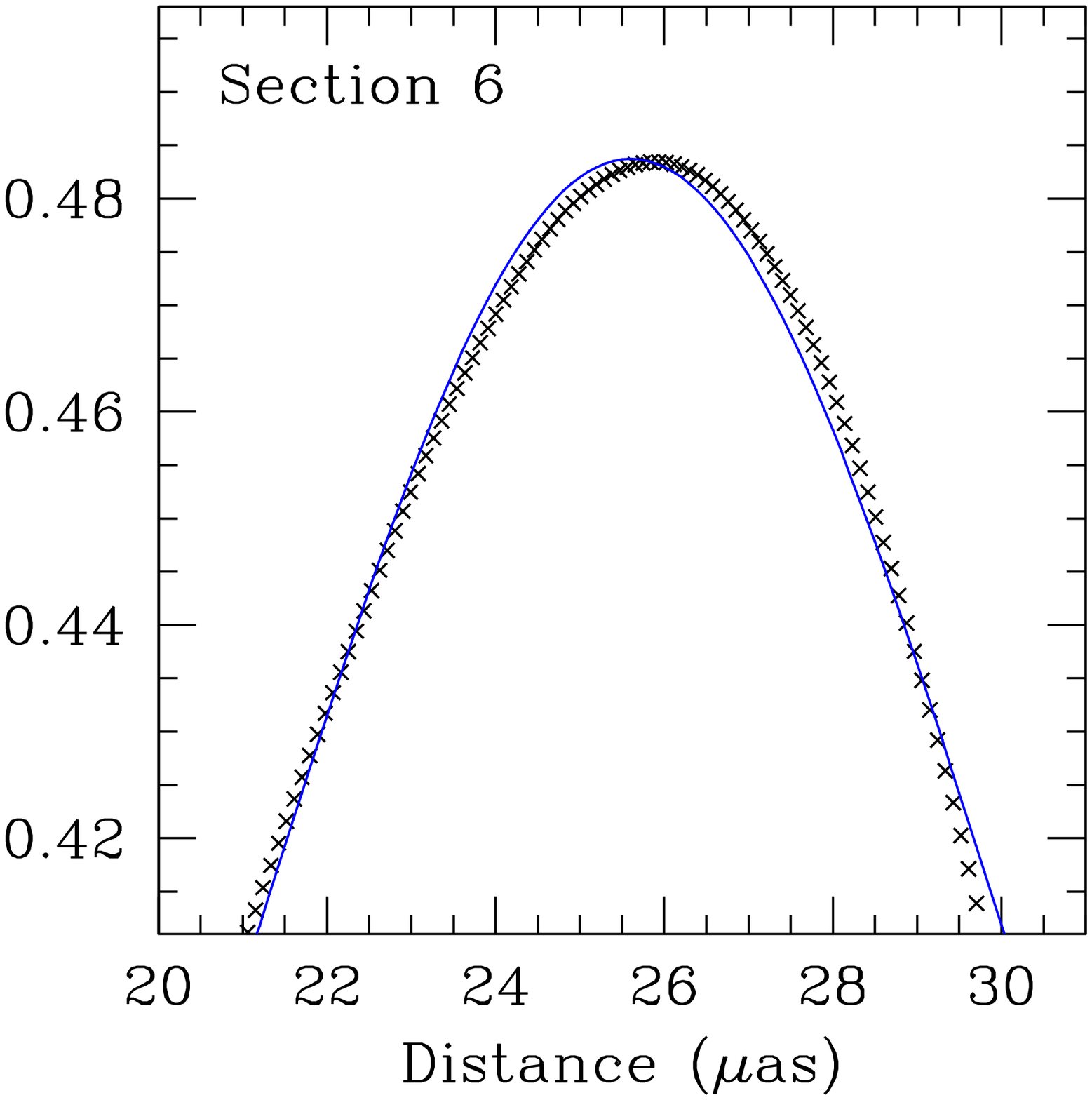,height=1.73in}
\psfig{figure=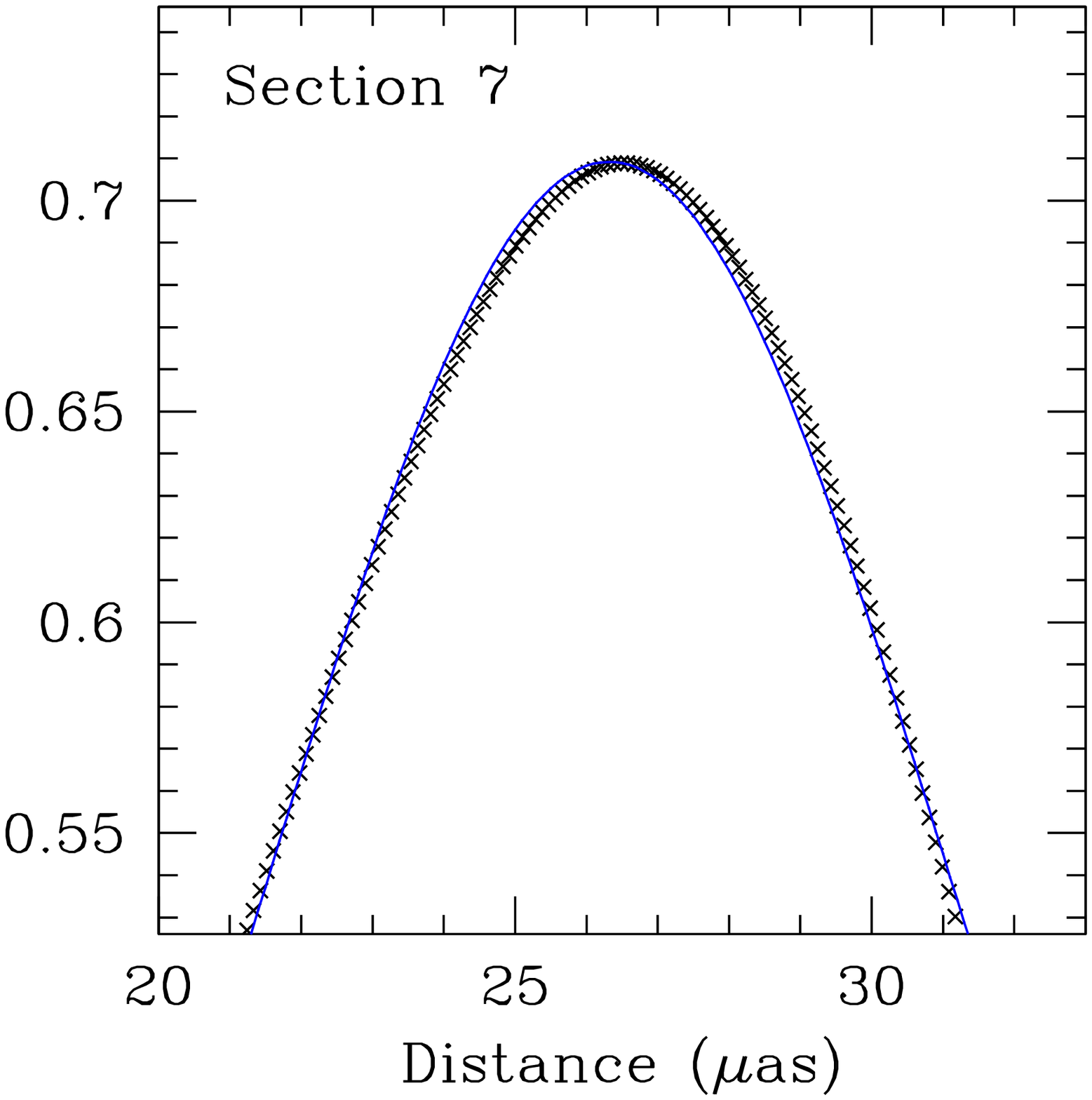,height=1.73in}
\psfig{figure=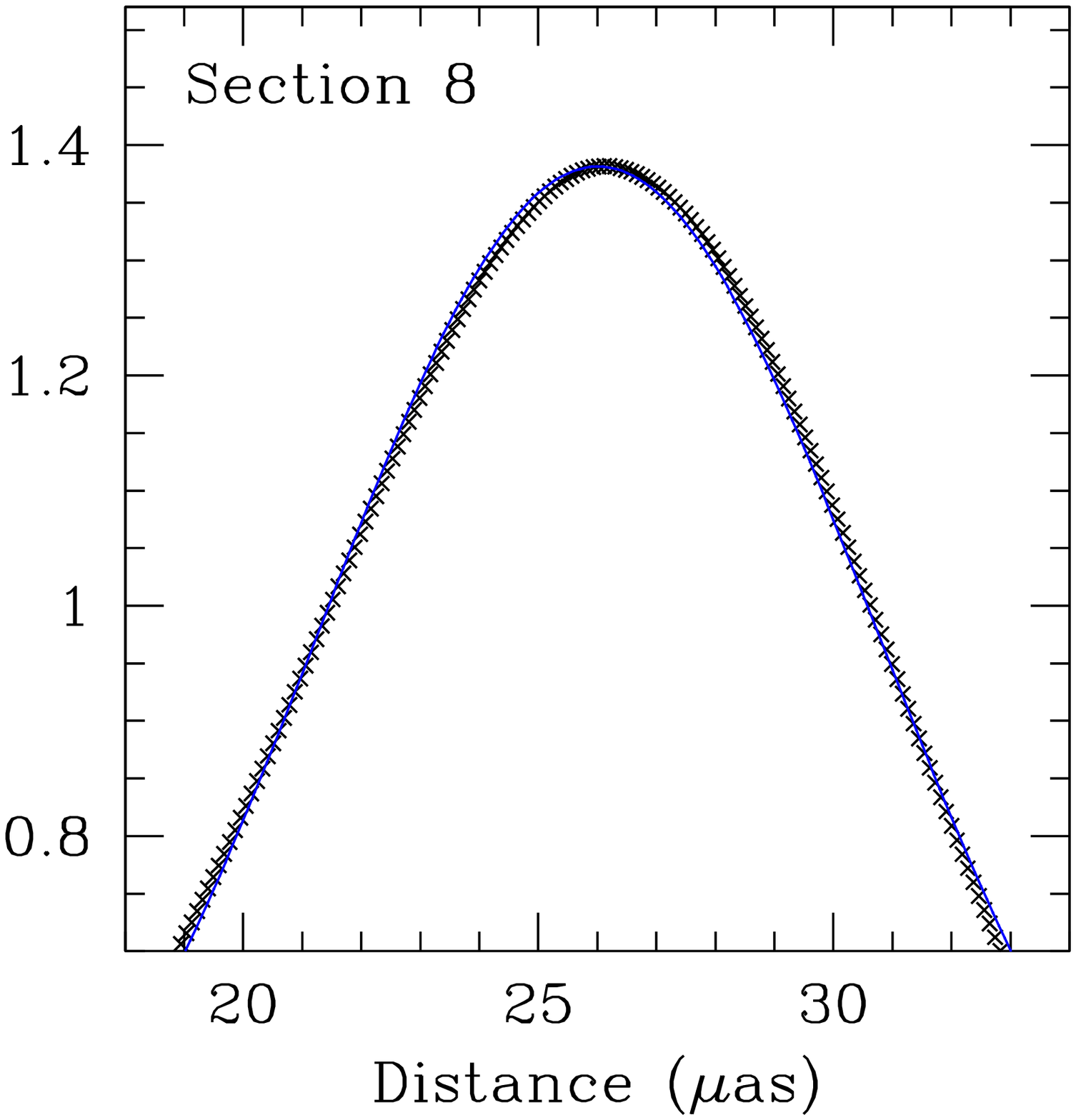,height=1.73in}
\end{center}
\caption{Gaussian fits (blue lines) of the specific intensity (black dots) along the chord sections 1-8 shown in the upper panel of Fig.~\ref{fig:ringimages}.}
\label{fig:gaussianfits}
\end{figure*}

{\it Image chord construction and measurement ---} We obtain the set of mean radii $\bar{r}_j$ along the chord sections $j=1,\ldots,8$ shown in the upper panel of Fig.~\ref{fig:ringimages} in the following manner. First, we draw a chord across the image at the approximate location of the spin axis of Sgr~A$^\ast$. We perform Gaussian fits of the specific intensity as a function of the angular radius $r_j$ along the sections of the image of the accretion flow labeled ``1'' and ``8'' and determine the mean angular radius and standard deviation of each Gaussian. We use the larger standard deviation of the two Gaussians to estimate the size of the effective resolution of the EHT, finding roughly $10~\mu{\rm as}$, which corresponds to an arc length of $\theta\approx22.3^\circ$ around the shadow. Next, we draw six additional equidistant chords across the image intersecting at the image center determined by the first chord, each separated by an angle $\theta=180^\circ/7\approx25.71^\circ$, and likewise perform Gaussian fits of the specific intensity along these chords at the sections labeled ``2'' to ``7.'' Since the specific intensity is enhanced on the ``bright'' side of the image due to relativistic beaming and boosting, which cause a partial obscuration of the shadow at 230~GHz, we focus only on the ``faint'' side of the image. Note that, even though the overall brightness of the image shown in Fig.~\ref{fig:ringimages} is asymmetric due to these relativistic effects, this part of the image has a nearly circular shape and the shape of the shadow itself is exactly circular. Fig.~\ref{fig:gaussianfits} shows the Gaussian fits of the specific intensity. The corresponding mean radii $\bar{r}_j$ and standard deviations $\bar{\sigma}_j$ of the Gaussian fits are listed in Table~\ref{tab:II}. From the set of angular radii $\bar{r}_j$ we then determine a suitable starting point for the subsequent Markov chain Monte Carlo sampling of a small region around the center of the chords from a least-squares fit of our model for the angular radii (given by the identity ${\bf\bar{r}_j}={\bf R}-{\bf x}$) based on a rough grid search.

\begin{table}[h]
\begin{center}
\footnotesize
\begin{tabular}{ccc}
\multicolumn{3}{c}{}\\
Section   & ~~$\bar{r}$  ($\mu$as)  & ~~$\bar{\sigma}$ ($\mu$as)  \\
\hline \hline
1	&23.81	&4.36 \\
2	&28.71	&2.95 \\
3	&27.40	&2.42 \\
4	&23.98	&1.60 \\
5	&22.72	&2.66 \\
6	&25.60	&4.27 \\
7	&26.32	&4.76 \\
8	&26.01	&4.73 \\
\hline
\end{tabular}
\caption{Fit parameters.}
\label{tab:II}
\end{center}
\end{table}

{\it Additional sources of uncertainty ---} In our estimate of the angular radius of the shadow we have also neglected two other sources of uncertainty. Our estimate will also be affected by calibration uncertainties of the noise level at individual EHT stations. EHT observations with a three-station array comprised by the James Clerk Maxwell Telescope (JCMT) and Sub-Millimeter Array (SMA) in Hawaii, the Submillimeter Telescope Observatory (SMTO) in Arizona, and several dishes of the Combined Array for Research in Millimeter-wave Astronomy (CARMA) in California have estimated 5\% systematic uncertainties from calibration for their visibility amplitudes~\cite{Fish11}. Our simulation is based on a seven-station EHT array, for which many more internal cross-checks will be available to improve the relative calibration of stations (the absolute calibration is not important). In particular, the use of three individual phased interferometers [Hawaii, CARMA, Atacama Large Millimeter/submillimeter Array (ALMA)] that simultaneously record conventional interferometric data will permit scan-by-scan cross calibration of the amplitude scale of the array. In addition, measurements of closure phases and closure amplitudes along different telescope triangles and quadrangles are immune to calibration errors. Repeating our analysis for a number of intentionally miscalibrated images will then allow us to estimate the uncertainty of the angular radius from the resulting distribution of angular radii obtained individually from each reconstructed image.

Another source of uncertainty is our current lack of knowledge of the detailed structure of the accretion flow of Sgr~A$^\ast$. Several accretion flow models have been proposed over the years~\cite{flowmodels}, many of which fall within the RIAF category. These models, in turn, depend on different parameters such as the vertical structure of the accretion disk, the plasma density and magnetic field strength in and above the disk, and the presence of collimated or uncollimated outflows such as a jet or winds. The full impact of these model uncertainties is a topic of ongoing research and will be analyzed elsewhere.

Our estimate does not apply to images within which greater parts of the shadow are obscured and, thus, less than eight chord sections would have to be used to measure the angular radius of the shadow. Although our method relies on the presence of an accretion flow which emits the radiation that comprises the bright ring surrounding the shadow, the brightness profiles along the different chords in the image will have local peaks near the location of the shadow irrespective of the details of the accretion flow itself (c.f., Ref.~\cite{pathlength}).

This ring is clearly seen in different general-relativistic magnetohydrodynamic simulations reported to date~\cite{GRMHD}; see, also, Fig.~8 in Ref.~\cite{PJproc}. We use a physically reasonable RIAF model which has provided consistent fits to EHT data collected over seven years~\cite{BroRIAF}. As such, any observed image will be immediately diagnostic of the applicability of the assumed accretion flow model. Nonetheless, our results should be viewed with caution until the uncertainties regarding the structure of the accretion flow are fully assessed. These systematic uncertainties will then have to be incorporated into our estimate of the angular radius of the shadow in a manner that is similar to the inclusion of the calibration uncertainties discussed above.

The improvement of EHT observations in the $N=100$ case together with stellar-orbit observations using a 30-m-class optical telescope compared to the $N=10$ case together with existing stellar-orbits data is about a factor of two. This improvement is primarily hampered by the uncertainty of the spin and the inclination of Sgr A$^\ast$ for which we simply assume a flat and an isotropic distribution, respectively, so that our estimate remains largely model-independent. In practice, long-term monitoring of stellar orbits will most likely lead to an independent spin measurement (e.g.,~\cite{WillApJL,GRAVITY}). Other studies, such as observations of the accretion flow (e.g.,~\cite{Bro14}) and of variability (e.g.,~\cite{QPOs}), can produce additional measurements of the spin and the inclination which would reduce that uncertainty considerably.

Interpreting combined data sets must be done with great care due to the difficulty in properly including their independent systematic uncertainties, which are likely to dominate their error budgets. For example, the current orbit-based measurements of Refs.~\cite{Gillessen09,Meyer12} nearly disagree at a statistically-significant level and Ref.~\cite{Reid14} neglects systematic errors arising from their choice of outlier removal and possible deviations from an axisymmetric velocity field.

Choosing different input values of the spin and inclination of Sgr~A$^\ast$ in our simulation of the constraints on its mass, distance, and deviation parameters, we find that the locations of the peaks of the distributions shown in Fig.~\ref{fig:constraints} depend on the corresponding shadow radius. However, the peak locations always lie within the $1\sigma$ confidence limits of the corresponding distributions, except for the peaks of the distributions of the deviation parameters in the $N=100$ case for values of the spin $a\gtrsim0.6r_g$ (regardless of the inclination). At that level of precision, the systematic uncertainty of the parameters $\alpha_{13}$ and $\beta$ caused by the dependence of the shadow size on the spin and inclination exceeds the corresponding $1\sigma$ confidence limits.


\begin{thebibliography}{10}
\makeatletter

\bibitem{Ghez08}
A.M. Ghez, S. Salim, N.N. Weinberg, {\it et al.}, Astrophys. J. {\bf 689}, 1044 (2008).

\bibitem{Gillessen09}
S. Gillessen, F. Eisenhauer, S. Trippe, {\it et al.}, Astrophys. J. {\bf 692}, 1075 (2009).

\bibitem{Meyer12}
L. Meyer, A.M. Ghez, R. Sch\"odel, S. Yelda, A. Boehle, J.R. Lu, T. Do, M.R. Morris, E.E. Becklin, and K. Matthews, Science {\bf 338}, 84 (2012).

\bibitem{Do13}
T. Do, G.D. Martinez, S. Yelda, A.M. Ghez, J. Bullock, M. Kaplinghat, J.R. Lu, A.G.H. Peter, and K. Phifer, Astrophys. J. {\bf 779}, L6 (2013).

\bibitem{Chatzopoulos15}
S. Chatzopoulos, T.K. Fritz, O. Gerhard, S. Gillessen, C. Wegg, R. Genzel, and O. Pfuhl, Mon. Not. R. Astron. Soc. {\bf 447}, 948 (2015).

\bibitem{Reid14}
M.J. Reid, K.M. Menten, A. Brunthaler, {\it et al.}, Astrophys. J. {\bf 783}, 130 (2014).

\bibitem{Doele08}
S.S. Doeleman, J. Weintroub, A.E.E. Rogers, {\it et al.}, Nature (London) {\bf 455}, 78 (2008).

\bibitem{heu96} 
M. Heusler, {\it Black Hole Uniqueness Theorems} (Cambridge University Press, Cambridge, England, 1996).

\bibitem{psalLRR} 
D. Psaltis, Living Rev. Relativity {\bf 11}, 9 (2008).

\bibitem{J13metric}
T. Johannsen, Phys. Rev. D {\bf 88}, 044002 (2013).

\bibitem{Supplementary}
See Supplemental Material, which includes Refs.~\cite{GB06,VH10,VYS11,JPmetric,Wald84,pathologies,MTW,WillLRR,Will93,YYT12,Jreview,Bro14,GairLRR,QPOs,Liu12,Xrayprobes,Fish11,flowmodels,GRMHD,PJproc,BroRIAF,WillApJL}, for the generality of Kerr-like metrics, implications for alternative gravity theories, other EHT gravity tests, image chord construction and measurement, as well as additional sources of uncertainty.

\bibitem{GB06}
K. Glampedakis and S. Babak, Class. Quantum Grav. {\bf 23}, 4167 (2006).

\bibitem{VH10}
S.J. Vigeland and S.A. Hughes, Phys. Rev. D {\bf 81}, 024030 (2010).

\bibitem{VYS11}
S.J. Vigeland, N. Yunes, and L.C. Stein, Phys. Rev. D {\bf 83}, 104027 (2011).

\bibitem{JPmetric}
T. Johannsen and D. Psaltis, Phys. Rev. D {\bf 83}, 124015 (2011).

\bibitem{Wald84}
R.M. Wald, {\it General Relativity} (University of Chicago Press, Chicago, 1984).

\bibitem{pathologies}
T. Johannsen, Phys. Rev. D {\bf 87}, 124010 (2013).

\bibitem{MTW}
C.W. Misner, K.S. Thorne, and J.A. Wheeler, {\it Gravitation} (W. H. Freeman and Co., New York, 1973).

\bibitem{WillLRR}
C.M. Will, Living Rev. Relativity {\bf 17}, 4 (2014).

\bibitem{Will93}
C.M. Will, {\it Theory and Experiment in Gravitational Physics} (Cambridge University Press, Cambridge, England, 1993).

\bibitem{YYT12}
K. Yagi, N. Yunes, and T. Tanaka, Phys. Rev. D {\bf 86}, 044037 (2012).

\bibitem{Jreview}
T. Johannsen, arXiv:1512.03818.

\bibitem{Bro14}
A.E. Broderick, T. Johannsen, A. Loeb, and D. Psaltis, Astrophys. J. {\bf 784}, 7 (2014).

\bibitem{GairLRR}
J.R. Gair, M. Vallisneri, S.L. Larson, and J.G. Baker, Living Rev. Relativity {\bf 16}, 7 (2013).

\bibitem{QPOs}
B. Aschenbach, N. Grosso, D. Porquet, and P. Predehl, Astron. Astrophys. {\bf 417}, 71 (2004); S. Trippe, T. Paumard, T. Ott, S. Gillessen, F. Eisenhauer, F. Martins, and R. Genzel, Mon. Not. R. Astron. Soc. {\bf 375}, 764 (2007).

\bibitem{Liu12}
K. Liu, N. Wex, M. Kramer, J.M. Cordes, and T.J.W. Lazio , Astrophys. J. {\bf 747}, 1 (2012).

\bibitem{Xrayprobes}
T. Johannsen, Phys. Rev. D {\bf 90}< 064002 (2014); N. Lin, Z. Li, J. Arthur, R. Asquith, and C. Bambi, J. Cosmol. Astropart. Phys. {\bf 09} (2015) 038.

\bibitem{Fish11}
V.L. Fish, S.S. Doeleman, C. Beaudoin, {\it et al.}, Astrophys. J. {\bf 727} L36 (2011).

\bibitem{flowmodels}
R. Narayan, R. Mahadevan, J E. Grindlay, R.G. Popham, and C.F. Gammie, Astrophys. J. {\bf 492}, 554 (1998); R.D. Blandford and M.C. Begelman, Mon. Not. R. Astron. Soc. {\bf 303}, L1 (1999); H. Falcke and P.L. Biermann, Astron. Astrophys. {\bf 342}, 49 (1999); F. \"Ozel, D. Psaltis, and R. Narayan, Astrophys. J. {\bf 541}, 234 (2000); F. Yuan, E. Quataert, and R. Narayan, Astrophys. J. {\bf 598}, 301 (2003).

\bibitem{GRMHD}
M. Mo\'scibrodzka, C.F. Gammie, J.C. Dolence, H. Shiokawa, and P.K. Leung, Astrophys. J. {\bf 706}, 497 (2009); J. Dexter, E. Agol, and P.C. Fragile, Astrophys. J. {\bf 703}, L142 (2009); R. Shcherbakov and R. Penna, in {\it The Galactic Center: A Window to the Nuclear Activity of Disk Galaxies}, edited by M. Morris {\it et al.} (Astronomical Society of the Pacific, San Francisco, 2011), p. 372; C.-K. Chan, D. Psaltis, F. \"Ozel, R. Narayan, and A. Sadowski, Astrophys. J. {\bf 799}, 1 (2015).

\bibitem{PJproc}
D. Psaltis and T. Johannsen, J. Phys. Conf. Ser. {\bf 283}, 012030 (2011).

\bibitem{BroRIAF}
A.E. Broderick, V.L. Fish, S.S. Doeleman, and A. Loeb, Astrophys. J. {\bf 697}, 45 (2009); {\bf 735}, 110 (2011); A.E. Broderick, V.L. Fish, M.D. Johnson, {\it et al.} (unpublished).

\bibitem{WillApJL}
C.M. Will, Astrophys. J. {\bf 674}, L25 (2008).

\bibitem{EHT}
S.S. Doeleman, E. Agol, D. Backer, {\it et al.}, {\it Astro2010: The Astronomy and Astrophysics Decadal Survey}, Science White Papers Vol. 68 (National Academies of Sciences, Engineering, and Medicine, Board on Physics and Astronomy, Washington, DC, 2009).

\bibitem{shadow}
J.M. Bardeen in {\it Black Holes}, (Gordon and Breach, New York, 1973); J.-P. Luminet, Astron. Astrophys. {\bf 75}, 228 (1979); H. Falcke, F. Melia, and E. Agol, Astrophys. J. {\bf 528}, L13 (2000); R. Takahashi, Astrophys. J. {\bf 611}, 996 (2004).

\bibitem{PaperII}
T. Johannsen and D. Psaltis, Astrophys. J. {\bf 718}, 446 (2010).

\bibitem{AE12}
L. Amarilla and E.F. Eiroa, Phys. Rev. D {\bf 85}, 064019 (2012).

\bibitem{J13rings} 
T. Johannsen, Astrophys. J. {\bf 777}, 170 (2013).

\bibitem{PsaltisNullTest}
D. Psaltis, F. \"Ozel, C.-K. Chan, and D.P. Marrone, Astrophys. J. {\bf 814}, 115 (2015).

\bibitem{SMBHmasses}
T. Johannsen, D. Psaltis, S. Gillessen, D.P. Marrone, F. \"Ozel, S.S Doeleman, and V.L. Fish, Astrophys. J. {\bf 758}, 30 (2012).

\bibitem{deblurring}
V.L. Fish, M.D. Johnson, R.-S. Lu, {\it et al.}, Astrophys. J. {\bf 795}, 134 (2014).

\bibitem{pathlength}
The specific intensity peaks at the shadow corresponding to the longest optical photon path in the accretion flow.

\bibitem{Weinberg95}
N.N. Weinberg, M. Milosavljevi\'c, and A.M. Ghez, Astrophys. J. {\bf 622}, 878 (2005).

\bibitem{RS2BH}
A.N. Aliev and A.E. G\"umr\"uk\c{c}\"uo\u{g}lu, Phys. Rev. D {\bf 71}, 104027 (2005).

\bibitem{MOG}
J.W. Moffat, Eur. Phys. J. C {\bf 75}, 175 (2015).

\bibitem{EdGB}
P. Pani, C.F.B. Macedo, L.C.B. Crispino, and V. Cardoso, Phys. Rev. D {\bf 84}, 087501 (2011); D. Ayzenberg and N. Yunes, Phys. Rev. D {\bf 90}, 044066 (2014); A. Maselli, P. Pani, L. Gualtieri, and V. Ferrari, Phys. Rev. D {\bf 92}, 083014 (2015).

\bibitem{Johnson15}
M.D. Johnson and C.R. Gwinn, Astrophys. J. {\bf 805}, 180 (2015).

\bibitem{Lu15}
R.-S. Lu, F. Roelofs, V.L. Fish, {\it et al.}, arXiv:1512.08543.

\bibitem{GRAVITY}
F. Eisenhauer, G. Perrin, W. Brandner, {\it et al.}, The Messenger {\bf 143}, 16 (2011).


\end{thebibliography}
\end{document}